\documentclass{ws-procs9x6}

\def\frac#1#2{{\textstyle{{#1}\over {#2}}}}

\def\lsim{\mathrel{\rlap{\lower4pt\hbox{\hskip1pt$\sim$}}
  \raise1pt\hbox{$<$}}}
\def\gsim{\mathrel{\rlap{\lower4pt\hbox{\hskip1pt$\sim$}}
  \raise1pt\hbox{$>$}}}
\def\sqr#1#2{{\vcenter{\vbox{\hrule height.#2pt
       \hbox{\vrule width.#2pt height#1pt \kern#1pt
       \vrule width.#2pt}
       \hrule height.#2pt}}}}

\newcommand{\beq}{\begin{equation}}
\newcommand{\eeq}{\end{equation}}
\newcommand{\bea}{\begin{eqnarray}}
\newcommand{\eea}{\end{eqnarray}}
\newcommand{\bit}{\begin{itemize}}
\newcommand{\eit}{\end{itemize}}

\begin{document}

\title{LORENTZ VIOLATION AND GRAVITY}

\author{R.\ BLUHM}

\address{Department of Physics, Colby College\\
Waterville, ME 04901, USA\\
E-mail: rtbluhm@colby.edu}

\begin{abstract}
Gravitational theories with Lorentz violation must account
for a number of possible features in order to be consistent
theoretically and phenomenologically.
A brief summary of these features is given here.
They include evasion of a no-go theorem,  
connections between spontaneous Lorentz breaking
and diffeomorphism breaking,
the appearance of massless Nambu-Goldstone modes
and massive Higgs modes,
and the possibility of a Higgs mechanism in gravity.
\end{abstract}

\bodymatter

\section{Gravity and the SME}	

The Standard-Model Extension\cite{sme} 
(SME) consists of the
most general observer-independent effective field theory
incorporating Lorentz violation.
It is routinely used by both theorists and experimentalists
to study and obtain bounds on possible 
forms of Lorentz violation.\cite{smereview,data}
As an effective field theory,
the SME can accommodate both explicit and spontaneous
Lorentz breaking.
However, there are differences in these two forms of 
symmetry breaking that arise in the context of gravity.
This overview looks at these differences and what their
primary consequences are.

In a gravitational theory with Lorentz violation it is useful
to use a vierbein formalism.  
In this approach, both the local Lorentz frames and
spacetime frames are accessible and linkage
between the symmetries in these frames can be examined.
The vierbein provides the
connection between tensor components in local Lorentz frames
and tensor components in the spacetime frame.

The Lagrangian in the SME is formed as the most general
scalar function (under both local Lorentz and diffeomorphism
transformations) using gravitational fields, particle fields, 
and Lorentz-violating SME coefficients.
When the Lorentz breaking is explicit,
the SME coefficients are viewed as fixed background fields.
However, when the Lorentz breaking is spontaneous, 
the SME coefficients are vacuum
expectation values (vevs) of dynamical tensor fields.

In the gravity sector of the SME,
a no-go theorem shows that with explicit Lorentz breaking
an inconsistency can occur between conditions stemming from
the field variations and symmetry considerations 
with geometrical constraints that must hold, 
such as the Bianchi identities.\cite{ak}
In contrast,
the case of spontaneous Lorentz breaking was
found to evade the no-go theorem.  
The main difference is that in a theory with explicit breaking
the SME coefficients are not associated with dynamical fields,
while with spontaneous Lorentz breaking they are,
which creates a difference in the conditions that must hold.
An important consequence of the no-go theorem is that the gravity sector
of the SME can only avoid incompatibility with conventional geometrical 
constraints if the symmetry breaking is spontaneous.

\section{Spontaneous symmetry breaking}

The fact that the SME coefficients must be associated with
vevs of dynamical fields that undergo spontaneous Lorentz
violation leads to a number of effects that must be accounted for
in the gravity sector of the SME.
For example, when Lorentz symmetry is spontaneously broken,
there is also spontaneous breaking of diffeomorphism symmetry.
The spontaneous Lorentz breaking occurs when
a nonzero tensor-valued vacuum occurs in the local Lorentz frames,
which is necessarily accompanied by a vacuum value for the vierbein. 
When products of the vierbein vev act on
the local tensor vevs, 
the result is that tensor vevs also appear in
the spacetime frame.
These tensor vevs spontaneously
break local diffeomorphisms in the spacetime frame.
Conversely, if a vev in the spacetime frame spontaneously
breaks diffeomorphisms,
then the inverse vierbein acting on it gives rise to vevs
in the local frames.
Consequently, spontaneous local Lorentz
breaking implies spontaneous diffeomorphism breaking
and vice versa.\cite{rbak}

With spontaneous symmetry breaking of both Lorentz and diffeomorphism symmetry,
there are standard features in particle physics that need to be investigated.
These include the possible appearance of massless Nambu-Goldstone (NG) modes
and massive Higgs modes, or there is the possibility of a Higgs mechanism
in which the NG modes are reinterpreted as additional degrees of freedom
in a theory with massive gauge fields.

In the absence of a Higgs mechanism,
there can be up to as many NG modes as there are broken spacetime symmetries. 
Since the maximal symmetry-breaking case would yield six broken Lorentz generators 
and four broken diffeomorphisms,
there can be as many as ten NG modes.
A natural gauge choice puts all of the NG modes into the vierbein.
However, 
this will in general lead to the appearance of ghosts,
and it is for this reason that most models involve breaking 
fewer than ten of the spacetime symmetries.

Spontaneous symmetry breaking is usually induced by a potential term in
the Lagrangian that has a degenerate minimum space.
The NG modes appear as excitations away from the vacuum
that stay in the minimum space,
while massive Higgs modes are excitations that go
up the potential well away from the minimum.  
In conventional gauge theory,
the potential involves only scalar fields,
and the massive Higgs modes are
independent of the gauge fields.
However, with spontaneous Lorentz breaking,
the metric typically appears in the potential along with the tensor fields,
and for this reason massive Higgs modes can occur
that include metric excitations.
This is an effect that has no
analog in the case of conventional gauge theory.

In a Higgs mechanism,
the would-be NG modes become additional degrees 
of freedom for massive gauge fields.
The gauge fields associated with diffeomorphisms 
are the metric excitations.
However,
a Higgs mechanism involving the metric 
has been shown not to occur.\cite{ks}  
This is because the mass term that is generated by covariant derivatives
involves the connection,  which consists of derivatives of the metric
and not the metric itself.
However, for the broken Lorentz symmetry, 
where the relevant gauge fields are the spin connection,
a conventional Higgs mechanism can occur.\cite{rbak}
This is because the spin connection appears directly in covariant derivatives acting 
on local tensor components,
and when the local tensors acquire a vev,
quadratic mass terms for the spin connection can be generated.
Note, however, a viable Higgs mechanism involving the spin connection can 
only occur if the spin connection is a dynamical field,
which requires nonzero torsion and that the geometry be Riemann-Cartan.

\end{document}